\documentclass[11pt,a4paper]{article}
\usepackage[utf8]{inputenc}
\usepackage[T1]{fontenc}
\usepackage[spanish]{babel}
\usepackage{comment}
\usepackage{amsmath}
\usepackage{amsfonts}
\usepackage{dirtytalk}
\usepackage{amssymb}
\usepackage[]{apacite}
\usepackage{url}
\usepackage{times}
\usepackage{graphicx}
\usepackage{float}
\graphicspath{ {images/} }
\usepackage{subcaption}
\usepackage{threeparttable}

\title{Predicción de la inflación en Costa Rica}
\author{Daniel Aguilar, Minor Acuña y Breyner Chacón }
\date{Noviembre 2023}

\begin{document}

\bibliographystyle{apacite}

\maketitle

\section{Introducci\'on}
La estabilidad de la tasa de inflaci\'on es una condici\'on necesaria para el buen funcionamiento de toda econom\'ia capitalista. En un ambiente econ\'omico con inflaci\'on vol\'atil, el crecimiento de la econom\'ia y la repartici\'on de este entre los agentes de la sociedad se ve comprometida. Por ello, y por cuanto en Costa Rica, como en la mayor parte de las naciones capitalistas, la autoridad monetaria est\'a encargada de mantener la estabilidad de precios, es necesario contar con modelos capaces de predecir el comportamiento de la inflaci\'on en el pa\'is. En esta l\'inea, en este trabajo se pretende comparar dos de los principales tipos de modelos predictivos b\'asicos encontrados en la literatura: modelos univariados autoregresivos, ejemplificados por los modelos ARIMA, y modelos multivariados formulados con base en la teor\'ia econ\'omica, como la curva de Phillips.

\section{Marco te\'orico}
\subsection{La inflaci\'on en la teor\'ia econ\'omica moderna}
En la actualidad, toda discusi\'on sobre los determinantes y relaciones de la tasa de inflaci\'on se centra en un grupo de modelos crucial: las curvas de Phillips. En su versi\'on actual, la tasa de inflaci\'on depende positivamente de dos factores fundamentales: alguna medida de la actividad econ\'omica real y las expectativas de inflaci\'on. La conjunci\'on de estos dos factores como determinantes del fen\'omeno inflacionario se puede comprender como el resultado de un proceso hist\'orico en la evoluci\'on de la teor\'ia econ\'omica. Por ello en esta secci\'on se expone primero una breve s\'intesis de la evoluci\'on de las teor\'ias de inflaci\'on para luego exponer la teor\'ia vigente. 

\subsubsection{La curva de Phillips}
El primer modelo inflacionario adoptado ampliamente por la comunidad acad\'emica es el modelo presentado por \cite{Phillips}. En este la inflaci\'on $\pi_t$ depende negativamente de la tasa de desempleo $u_t$ de la forma:
\begin{equation}\label{pc}
    \pi_t=\lambda u_t^{-\alpha} +\beta_0; \ \lambda, \alpha>0
\end{equation}
Esta relaci\'on se atribuy\'o a la idea de que, a mayor exceso de oferta en el mercado laboral, i.e., a mayor desempleo, mayor deb\'ia ser el valor absoluto del cambio porcentual en los salarios y, en la medida en la que estos influencian los precios, menor deb\'ia ser la tasa de inflaci\'on. El caracter no lineal de (\ref{pc}) se atribu\'ia a la rigidez a la baja de los salarios nominales\footnote{Este era un fen\'omeno cuya recurrencia era aceptada.}: aumentos en la tasa de desempleo ten\'ian un menor efecto sobre los salarios en la medida en la que estos se \say{resist\'ian} a disminuir. Durante la d\'ecada de 1960 esta fue la teor\'ia predominante sobre inflaci\'on y \ref{pc} disfrut\'o de cierto \'exito emp\'irico.

\subsubsection{Expectativas y tasa de desempleo natural}
A finales de los 60's, la curva de Phillips sufri\'o una importante cr\'itica te\'orica planteda por \cite{Friedman}. La curva de Phillips de entonces implicaba la existencia de un \say{men\'u} de combinaciones entre inflaci\'on y desempleo a disposici\'on de los bancos centrales. Por ello, se argumentaba que la autoridad monetaria pod\'ia disminuir el desempleo v\'ia pol\'itica monetaria: aumentar la masa de dinero generar\'ia inflaci\'on y con ello, como implica (\ref{pc}), el desempleo disminuir\'ia \cite{gordon11}. Para \cite{Friedman} este no era el caso por cuanto la relaci\'on en (\ref{pc}) era inestable ya que omit\'ia el rol de expectativas de los agentes. Estos, al enfrentarse a altas tasas de inflaci\'on, cambiar\'ian sus expectativas y as\'i tambi\'en sus conductas de forma tal que la tasa de inflaci\'on efectiva se presionase al alza\footnote{Por ejemplo, aumentos en la expectativas de inflaci\'on har\'ian que los trabajadores exigiesen salarios m\'as altos, subiendo as\'i los costos de las empresas y con ello los precios. Adicionalmente, cualquier aumento en la demanda generado por la disminuci\'on de los tipos de inter\'es que acompa\~{n}an a una expansi\'on de la oferta monetaria ser\'ia ef\'imero al ver los agentes un alza en la inflaci\'on}. Esto implicaba que para todo momento $t$ exist\'ia una tasa de desempleo \say{natural} $u_t^N$ cuyo valor era \emph{independiente} de la inflaci\'on y oferta monetaria. \\
\ La inestabilidad de (\ref{pc}) atribuida a esta cr\'itica se vio emp\'iricamente validada en la d\'ecada de los 70's cuando, en un contexto de importantes aumentos en los precios del petr\'oleo, se dieron importantes alzas en la inflaci\'on en occidente acompa\~{n}adas de decrecimiento en la producci\'on y alto desempleo. Todo esto motiv\'o a que se extendiera la curva de Phillips a una versi\'on que puede caracterizarse de la siguiente forma:
\begin{equation} \label{PC_exp}
    \pi_t = \beta E_{t-1}\{\pi_t\} + \lambda (u_t-u_t^N) + \gamma \cdot z_t; \ \ \gamma, \lambda, \beta. 
\end{equation}
con $E_{t-1}\{\pi_t\}$ la inflaci\'on esperada en el periodo previo y $z_t$ un vector de variables representativas de shocks de oferta, (como el precio del petr\'oleo) y $theta$  un vector de par\'ametros. Como las expectativas se consideraban dependientes de la inflaci\'on ya observada por los agentes, usualmente se tomaba $E_{t-1}\{\pi_t\}$ como una combinaci\'on lineal de rezagos en $\pi_t$, particularmente en la modelaci\'on econom\'etrica. Adicionalmente se sol\'ian incluir rezagos en la brecha en la tasa de desempleo, la cual se pod\'ia sustituir por la brecha en el PIB. De esta manera se incorporaron los dos factores fundamentales ya mencionados: una medida de la acvtidad econ\'omica real, dada por $u_t-u_t^N$, y las expectativas. Este fue el modelo predominante en la d\'ecada de los 80's y 90's en c\'irculos acad\'emicos y modelaci\'on estad\'istica de los determinantes de la inflaci\'on. 

\subsubsection{Modelos de formaci\'on de precios de la nueva macroeconom\'ia keynesiana}
Si bien el modelo planteado en (\ref{PC_exp}) present\'o cierto \'exito emp\'irico desde su postulaci\'on hasta la actualidad, carece de un fundamento te\'orico basado en el comportamiento de los agentes individuales que habitan la econom\'ia, lo cual, siguiendo la influyente cr\'itica de \cite{lucas_crit}, se ha convertido en una exigencia dentro de la econom\'ia ortodoxa. Ello llev\'o al desarrollo de modelos te\'oricos que admit\'ian relaciones similares a (\ref{PC_exp}) deducidas de las decisiones de agentes racionales que toman decisiones \'optimas sujetos a las restricciones impuestas por su entorno econ\'omico. De esto surgieron teor\'ias sobre formaci\'on de precios que, incorporando rigideces al cambio de precios, permiten la deducci\'on de la llamana curva de Phillips de la Nueva Macroeconom\'ia Keynesiana (NKPC), la cual es parte del actual paradigma sobre fluctuaciones econ\'omicas. A fin de ilustrar esto, ac\'a se expone el modelo de formaci\'on de precios de Calvo, que es la opci\'on usual encontrada en la literatura. En este modelo la econom\'ia est\'a poblada por un continuo de firmas indexadas por $i \in [0,1]$ que se diferencian únicamente por el bien que producen y su historial de precios y que se enfrentan a una función de demanda isoelástica de la forma 
\begin{equation*}
    Y_{ti}^d=(P_{ti}/P_t\ )^\epsilon\ Y_t
\end{equation*}
donde $Y_{ti}^d$ es la demanda del bien $i$, $P_{it}$ su precio y $P_t :=(\int_0^1P_{ti}^{1-\epsilon}di)^{\epsilon-1}$ es un \'indice de precios. Cada periodo, solo una fracción $1-\theta$ aleatoria de las firmas puede cambiar su precio. Para cada firma, en cada periodo, la probabilidad de pertenecer a una u otra fracción es independiente de su historial de precios. Aquellas firmas que pueden cambiar el precio fijan un precio que optimiza una sumatoria de los valores presentes esperados de los beneficios futuros. De ello y de la condici\'on de equilibrio en el resto de la econom\'ia se obtiene la NKPC
\begin{equation}\label{NKPC}
    \pi_t=\beta E_t\{\pi_{t+1}\}+\lambda y_t
\end{equation}
donde $E_t\{\pi_{t+1}\}$ son las expectativas en $t$ respecto de $\pi_{t+1}$, $y_t$ es la desviaci\'on logar\'itmica del PIB respecto de su nivel natural\footnote{Es decir, $y_t=ln(Y_t)-ln(Y_t^N)$ con $Y_t^N$ el PIB natural}, $\beta>0$ es un factor de descuento \emph{sobre la utilidad de los hogares} y $\lambda=\theta^{-1}(1-\theta)(1-\beta \theta)>0$. De esta manera la inflaci\'on en $t$ depende (positivamente) de las expectativas que hoy se tienen respecto de la inflaci\'on futura y del exceso en la producci\'on agregada sobre las capacidades de la econom\'ia. 
La NKPC en (\ref{NKPC}) constituye el modelo b\'asico sobre inflaci\'on en la actual ortodoxia econ\'omica y se utiliza como punto de partida para posibles extensiones. Crucialmente, la NKPC contrasta con (\ref{PC_exp}) en el hecho de que son las expectativas \emph{actuales sobre la inflaci\'on presente} y no las expectativas previas sobre la inflaci\'on de hoy. 

\subsubsection{Extensiones a la NKPC}
Si bien la relaci\'on en (\ref{NKPC})  captura parte de la din\'amica inflacionaria, su \'exito no ha sido uniforme y la evidencia emp\'irica sugiere que el fen\'omeno es m\'as complejo que lo que la NKPC muestra. En particular, la literatura (ve\'ase \cite{ABBAS2016378}) sugiere dos principales extensiones a la NKPC. Primero, parte de la literatura considera la tasa de inflaci\'on como una variable con inercia en el sentido de que esta parece depender fundamentalmente de sus rezagos\footnote{De ah\'i el \'exito atribuido a modelos autoregresivos univariados sobre modelos multivariados \cite{stock}}. Por ello, se ha considerado incorporar rezagos de la tasa de inflaci\'on a (\ref{NKPC}) lo cual tambi\'en es motivado por la idea de que los agentes, al formar sus expectativas $E_t\{\pi_{t+1}\}$ toman en cuenta la inflaci\'on pasada. Segundo, la NKPC en (\ref{NKPC}) no incluye expl\'icitamente las presiones inflacionarias ejercidas por influencias externas, particularmente por fluctuaciones en el tipo de cambio y cambios en los precios de las importaciones. Adem\'as de esto, en la teor\'ia econ\'omica se suele extender (\ref{NKPC}) incluyendo variables representativas de \emph{shocks} de oferta como precios de petr\'oleo, como en (\ref{PC_exp}), tipos de inter\'es internacionales\footnote{A mayor tipo de inter\'es en el extranjero, menor demanda de moneda dom\'estica y, as\'i, mayor inflaci\'on} y condiciones de cr\'edito dom\'esticas. 

\section{Prediciendo la tasa de inflaci\'on}
Los m\'etodos estad\'isticos usados para proyectar la tasa inflaci\'on pueden clasificarse en dos grandes grupos: modelos autoregresivos univariados y modelos multivariados basados, al menos en parte, en teor\'ia econ\'omica. Los primeros consisten en modelos en los que la inflaci\'on presente depende \'unicamente de sus valores previos y, en ciertos casos, de t\'erminos de error estoc\'asticos; quiz\'as los modelos m\'as usuales encontrados en esta clase son los modelos ARIMA.  
\par Cada uno de estos plantea problemas particulares al momento de proyectar. Por un lado, los modelos autoregresivos poseen la ventaja de que permiten proyectar la inflaci\'on futura sin necesidad de proyectar de forma simult\'anea el comportamiento de alguna covariable que influye \emph{contempor\'aneamente} sobre la inflaci\'on, lo cual suele suceder en los modelos multivariados. Es estos \'ultimos se suelen crear modelos autoregresivos \say{subsidiarios} que se utilizan para obtener proyecciones de las covariables que luego se introducen en el modelo base; e.g. se proyecta la inflaci\'on con un modelo de la forma (\ref{NKPC}) donde se le introducen las proyecciones de la brecha del PIB y expectativas obtenidas v\'ia modelos ARIMA. Por otro lado, los modelos multivariados permiten incorporar posibles din\'amicas inflacionarias m\'as complejas donde aspectos como la actividad econ\'omica real, comercio internacional y expectativas inciden sobre la inflaci\'on. El consenso en la literatura es que los modelos univariados tienen, en promedio, un mejor desempe\~{n}o que los multivariados, si bien estos superan a los primeros en episodios de desequilibrio econ\'omico como la estanflaci\'on de los 70's y la recesi\'on de inicios de los 90's \cite{stock}.

\section{Marco Metodologíco} \label{metod}
\par En este trabajo se estiman dos grupos de modelos predictivos de la tasa de inflaci\'on a fin de evaluar el desempe\~{n}o de cada uno. Sin embargo, previo la estimaci\'on, se realiza un an\'alisis exploratorio de los datos a fin de caracterizar el comportamiento de las variables y detectar posibles relaciones entre estas. En el caso de la tasa de inflaci\'on, se ejecuta una prueba de cambio estructural en la serie, dados los cambios en la pol\'itica monetaria y estructura macroecon\'omica del pa\'is ocurridos a inicios del siglo. En particular, en enero del 2005, \say{la Junta Directiva del BCCR aprobó el proyecto estratégico `Esquema de meta explícita de inflación para Costa Rica', cuyo objetivo fue diseñar una estrategia orientada a preparar las condiciones para adoptar un régimen monetario que permitiera a la economía costarricense alcanzar la
estabilidad de precios en el mediano plazo} \cite{munoz2018adopcion}. Desde entonces el BCCR ha ido orientando su pol\'itica monetaria a mantener inflaci\'on baja y estable, culminando con la introducci\'on del esquema de metas de inflaci\'on en enero del 2018. 
\par Respecto de los modelos estimados, primero se estiman dos modelos ARIMA para la tasa de inflaci\'on intertrimestral, cada uno dado para elecciones de rezagos distintas. Por un lado, se estima un modelo cuya cantidad de rezagos se obtiene por el criterio AIC. En particular, se usa una selecci\'on de modelo ARIMA de mejores subconjuntos bajo AIC. Por otro, se estima un modelo cuyos rezagos se eligen observando las funciones de correlaci\'on y autocorrelaci\'on en la forma usual de la metodolog\'ia LJung-Box. En todos los casos se utilizan estimadores de m\'axima verosimilitud. 
\par Para los modelos, basados en teor\'ia econ\'omica, se elige un modelo de mejores subconjuntos utilizando como covariables la brecha en el PIB, las expectativas de inflaci\'on interanual\footnote{Estas son las expectativas de inflaci\'on estimadas por el Banco Central de Costa Rica (BCCR) a partir de la aplicaci\'on de teor\'ias de no arbitraje y los datos de las transacciones en el mercado de bonos dom\'estico}, tasa de inflaci\'ion intertrimestral de materias primas importadas y tipo de cambio promedio trimestral. Adem\'as, se incluyen como covariables cuatro rezagos a la brecha del producto y dos rezagos a la inflaci\'on importada y expectativas de inflaci\'on a fin de considerar la posibilidad de efectos rezagados de estas variables sobre el fen\'omeno inflacionario, tal y como indica la teor\'ia. Obtenido este modelo base, para cada covariable obtenida en el modelo se estima un modelo ARIMA con rezagos elegidos bajo el criterio AIC v\'ia mejores subconjutos, de los cuales se extraen proyecciones que luego se introducen en el modelo original, como es est\'andar en la literatura \cite{stock}. Luego, se hace un an\'alisis del desempe\~{n}o de cada uno dentro y fuera de la muestra.
\par En la estimaci\'on y elecci\'on de todos los modelos se usa una muestra con fecha final dada por el cuarto trimestre del 2021, lo cual brinda la posibilidad de ver su desempe\~{n}o en proyectar la infalci\'on durante el siguiente a\~{n}o. 


\section{An\'alisis exploratorio de datos}

\subsection{Cambios estructurales en la tasa de inflaci\'on}
\begin{figure}[H]
\centering
    \caption{\\[0.0001cm] \small \textbf{Tasa de inflación intertrimestral}}
    \centering
    \includegraphics[scale=0.5]{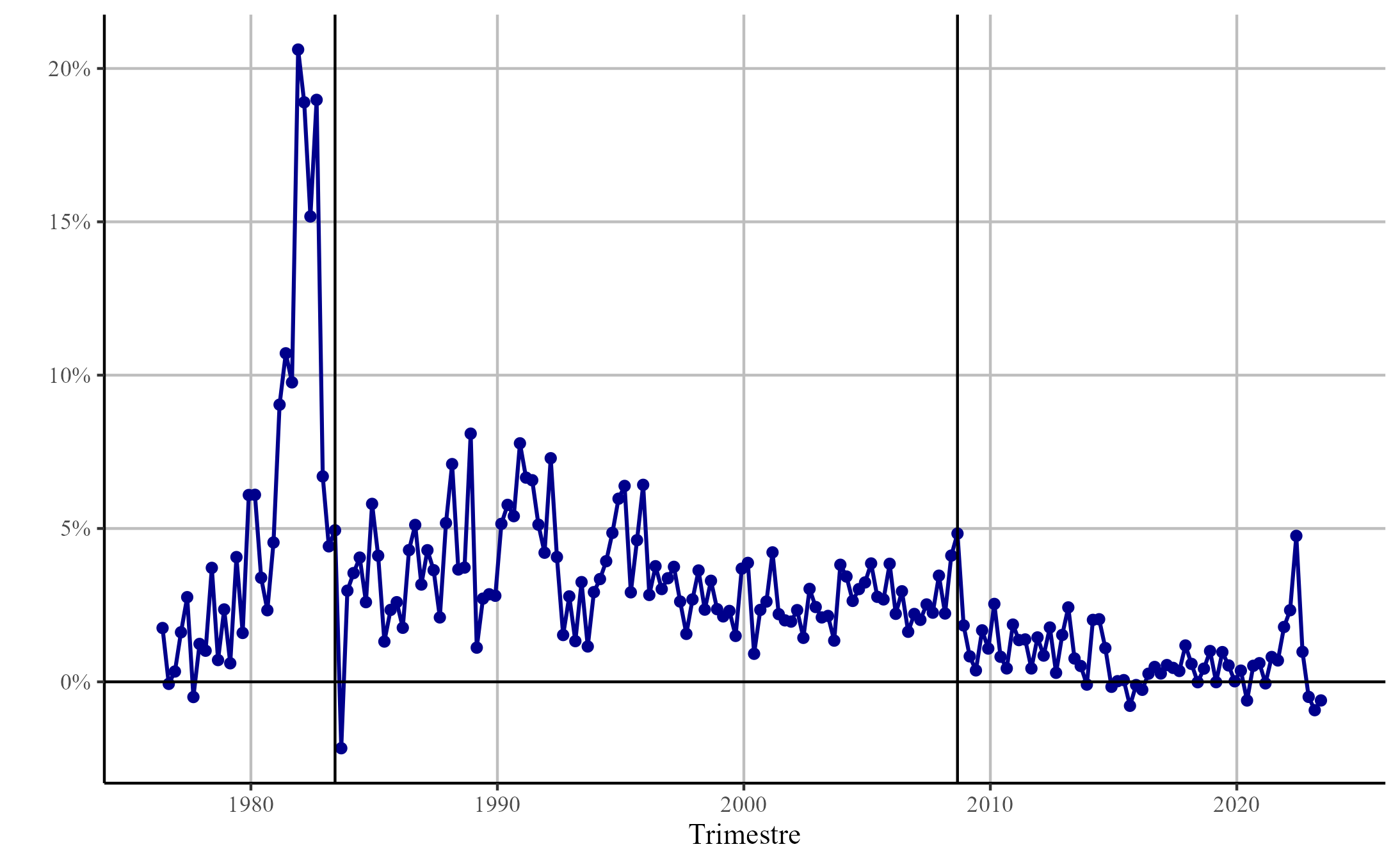}
    \begin{tablenotes}
\footnotesize
 \item \textbf{Fuente:} Elaboración propia .
    \end{tablenotes}
    \label{fig:inf_full}
\end{figure}

En la figura \ref{fig:inf_full} se muestra la evoluci\'on de la tasa de inflaci\'on intertrimestral para toda la muestra disponible. Destacan tres principales episodios. Primero, un episodio de hiper inflaci\'on a inicios de la muestra y hasta inicios de los 80's que conicide con la crisis econ\'omica de entonces. Segundo, un episodio de inflaci\'on m\'as moderada pero a\'un vol\'atil hasta mediados de la primer d\'ecada del siglo. Tercero, un episodio de inflaci\'on mucho m\'as baja y menos volatil que coincide con los primeros pasos tomados por el BCCR para la introducci\'on de un esquema de metas de inflaci\'on.

A fin de valorar la posibilidad de un cambio estructural en la inflaci\'on del pa\'is, lo cual ha sido documentado en la literatura (e.g. \cite{munoz2018adopcion}) se corri\'o una prueba de cambio estructural de minimizaci\'on de cuadrados residuales. Con ello se indentific\'o la presencia de dos puntos de quiebre en la serie: segundo trimestre de 1983 y primer trimestre del 2008, que corresponden a las rectas verticales en la figura \ref{fig:inf_full}. Dado este hallazgo y que el trabajo est\'a orientado a proyectar la tasa de inflaci\'on bajo el actual esquema de pol\'itica monetaria, se decide tomar la muestra a partir del primer trimestre del 2009,lo cual tambi\'en permite el uso de la serie de expectativas de inflaci\'on de mercado que no est\'a disponible para fechas previas, y hasta el \'ultimo trimestre del 2018, a fin de poder evaluar el desmepe\~{n}o de los modelos fuera del periodo de la muestra. 

\subsection{Medidas b\'asicas}
\begin{figure}[H]
\centering
    \caption{\\[0.0001cm] \small \textbf{Correlaci\'on de tasa de inflaci\'on intertrimestral con covaribales, 1Q2009-4Q2021}}
    \centering
    \includegraphics[scale=0.6]{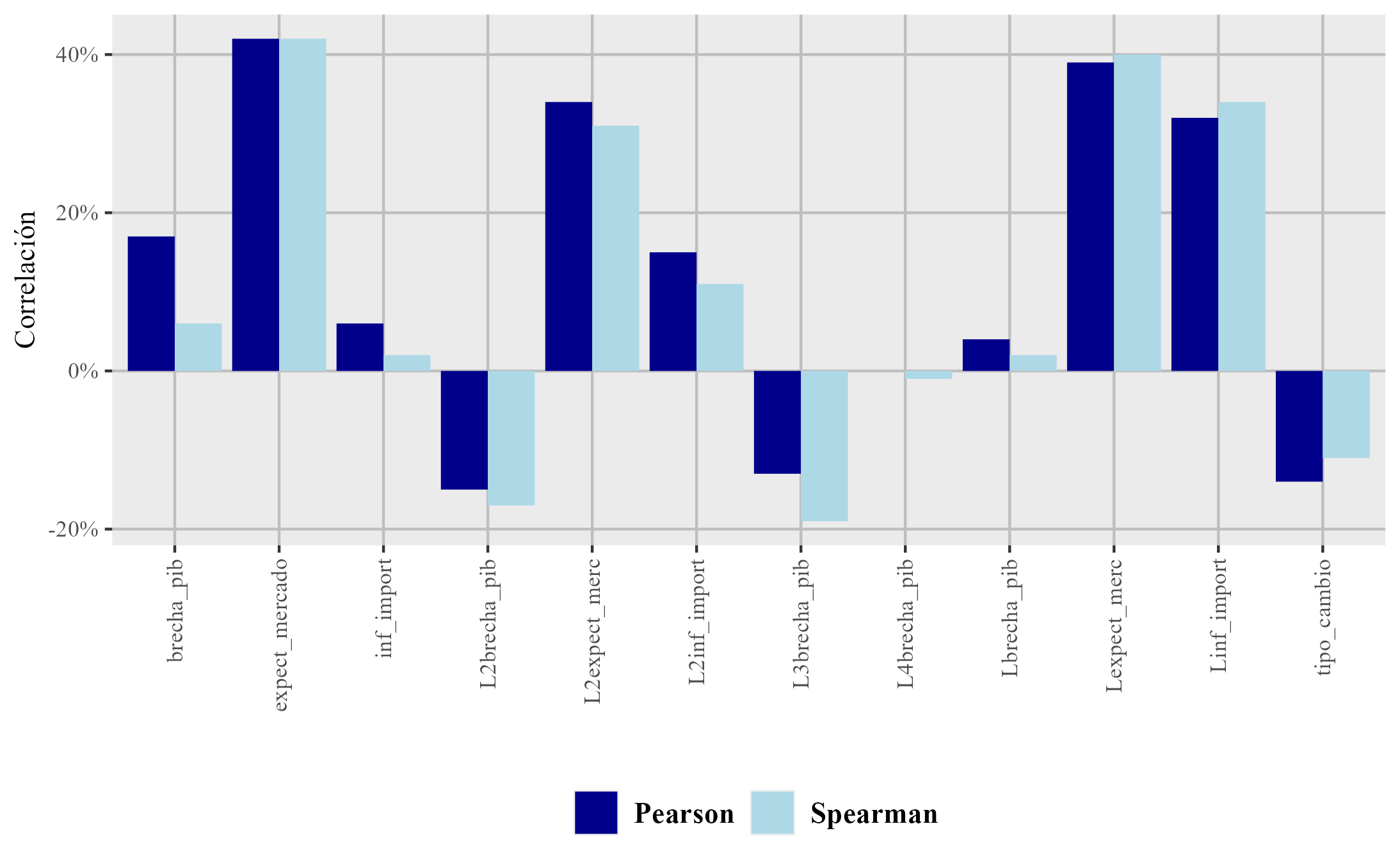}
    \label{fig:plot_corr_sub3}
\end{figure}

En esta secci\'on se muestran las medidas b\'asicas para las variables en la base para la muestra tomada. En la figura \ref{fig:plot_corr_sub3} se muestra la correlaci\'on de la tasa de inflaci\'on y el resto de variables covariables en la base. Las expectativas de inflaci\'on constituyen la variable con mayor correlaci\'on con un nivel superior al 50\%, tanto en su valor contempor\'aneo como en su segundo y tercer rezagos. En todos los casos la correlaci\'on es positiva lo que sugiere que las expectativas presionan la inflaci\'on al alza y viceversa. La inflaci\'on de materias primas importadas es la segunda covariable con mayor correlaci\'on, rondando el 30\%, lo que sugiere la existencia de una presi\'on al alza de la inflaci\'on proveniente de los mercados extranjeros. Por otro lado, la brecha del PIB presenta una muy baja correlaci\'on positiva en su valor contempor\'aneo que aumenta marginlamente en el primer rezago, mientras que, para el tercer y cuarto rezago, la correlaci\'on aumenta pero se vuelve negativa. El tipo de cambio presenta una correlaci\'on relativamente baja y negativa. 

En el cuadro \ref{tabla_meds} se muestran la media, \'inimos, m\'aximos y cuartiles de las variables en la muestra. Destaca el hecho que de el m\'nimo en la inflaci\'on es negativo, i.e.,  deflaci\'on, mientras que no es as\'i en las expecatativas pero s\'i en la brecha del PIB e inflaci\'on de materias primas importadas. 

\begin{table}[H]
\centering
\caption{\\[0.0001cm] \small \textbf{Media, m\'inimos, m\'aximos y cuartiles de variables, 1Q2009 a 4Q2018}}
\begin{tabular}{rlllll}
  \hline
 &    Inflaci\'on &   Inflaci\'on   & Expectativas &  Tipo de cambio &   Brecha PIB \\ 
 & & importadas& \\
  \hline
   M\'inimo &  -0.0078   &    -0.3304   &    0.0104   &    499.0   &    -0.1023   \\ 
  1er Quartil &  0.0026   & -0.0534   & 0.0224   & 511.1   & -0.0169   \\ 
  Mediana &  0.0054   & 0.0239   & 0.0315   &  550.4   &  0.0022   \\ 
  Media & 0.0073   &   0.0085   & 0.0294   &   550.3   &    -0.0029   \\ 
  3er Quartil &  0.0116   &  0.0701   & 0.0359   & 573.8   &  0.0169   \\ 
M\'aximo &     0.0254   &     0.2599   &    0.054   &    635.1   &     0.0420   \\ 
   \hline
\end{tabular}
\label{tabla_meds}
\end{table}

\begin{figure}[H]
\centering
\caption{\\[0.0001cm] \small \textbf{Autocorrelaci\'on y autocorrelaci\'on parcial de la tasa de inflaci\'on, 1Q2009 a 4Q2021}}
\begin{subfigure}[b]{0.47\textwidth}
    \centering
    \includegraphics[width=\textwidth]{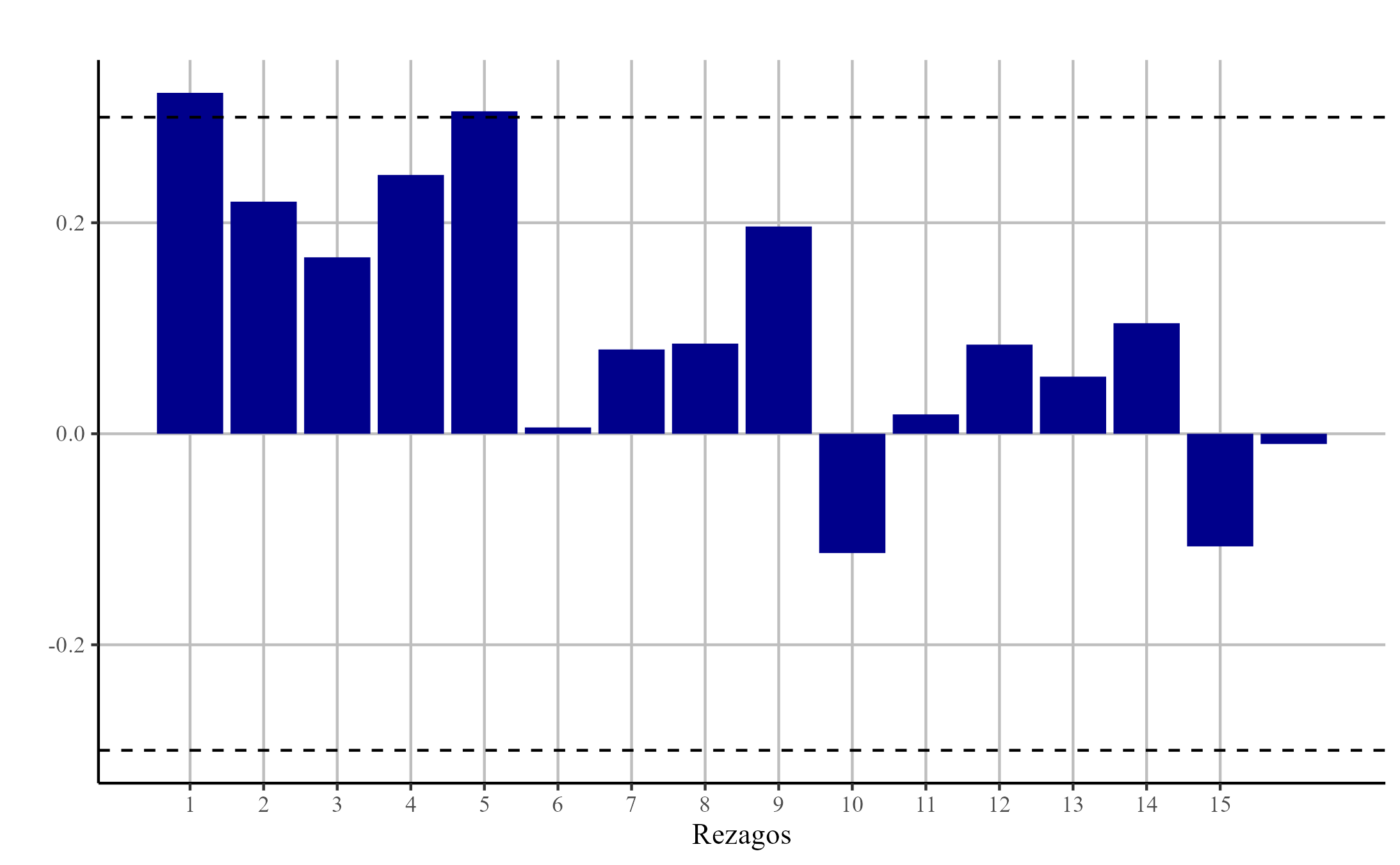}
    \caption{Autocorrelaci\'on}
\end{subfigure}
\hfill
\begin{subfigure} [b]{0.47\textwidth}
    \centering
    \includegraphics[width=\textwidth]{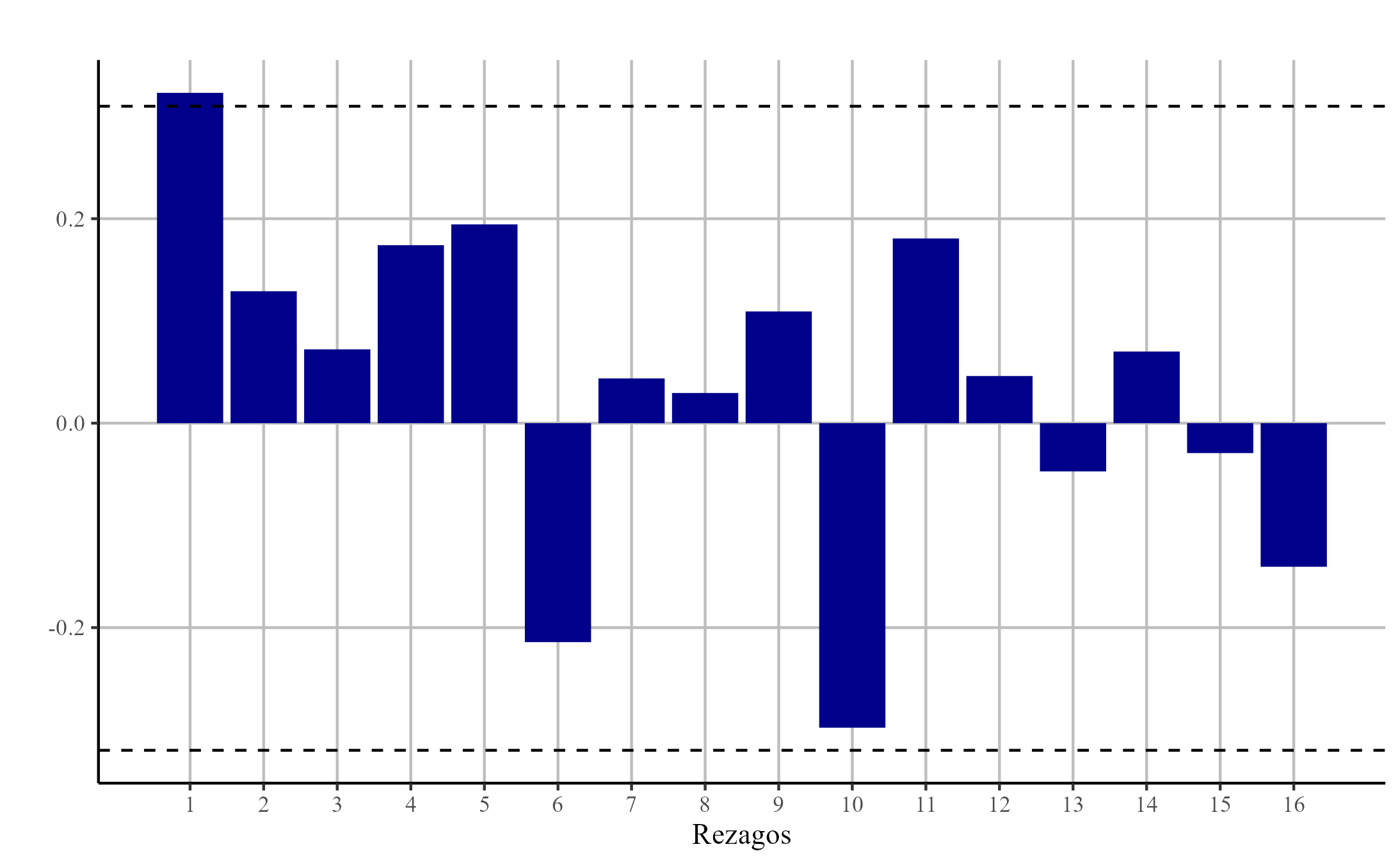}
    \caption{Autocorrelaci\'on parcial}
\end{subfigure}
\label{fig:afc_pacf}
\end{figure}

Finalmente, en la figura \ref{fig:afc_pacf} se muestran las las gr\'aficas con la autocorrelaci\'on y autocorrelaci\'on parcial de la tasa de inflaci\'on para la muestra tomada. Bajo la medida autocorrelaci\'on resultan significativos el primer y quinto rezagos, mientras que para la autocorrelaci\'on parcial solo el primero y el d\'ecimo. Esto sugiere un comportamiento relativamente poco inercial de la tasa de inflaci\'on, i,e., sus valores actuales no parecen dependender de forma fundamental de sus valores previos.  

\subsection{Evoluci\'on de variables}

\begin{figure}[H]
\centering
    \caption{\\[0.0001cm] \small \textbf{Tasa de inflación intertrimestral}}
    \centering
    \includegraphics[scale=0.5]{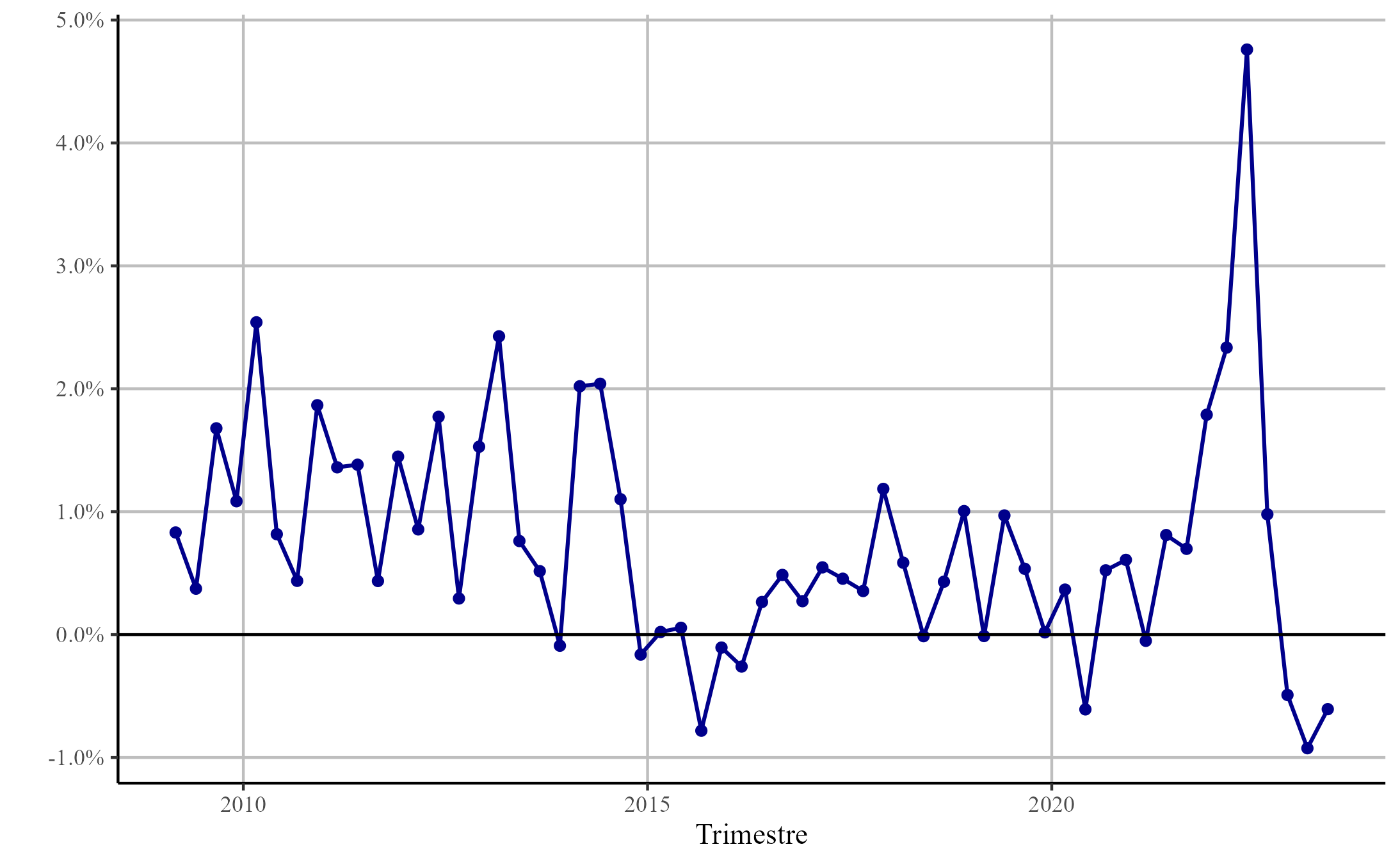}
    \begin{tablenotes}
\footnotesize
 \item \textbf{Fuente:} Elaboración propia .
    \end{tablenotes}
    \label{fig:plot_inf_q}
\end{figure}

En el grafico \ref{fig:plot_inf_q} se muestra la evoluci\'on de la tasa de inflaci\'on a partir del inicio de la muestra. Destacan tres aspectos. Primero,  se presentan varios trimestres con deflaci\'on alrededor del 2015 \footnote{En particular, el cuarto semestre del 2013 y 2014, el tercer y cuarto semestre del 2015 y primero del 2016, el segundo del 2018 y primero del 2019, presentan tasa de inflaci\'on negativa.} y luego del 2020, incluyendo los dos primeros trimestres del 2023. Segundo, se puede distinguir una disminuci\'on a\'un menor de la volatilidad de la inflaci\'on tras el 2015. Tercero, se observan picos en la inflaci\'on del \'ulitmo trimestre del 2021 y los dos primeros del 2022, los que podr\'ian ser causados por presiones externas dado el contexto global de entonces. 

\par
\begin{figure}[H]
\centering
    \caption{\\[0.0001cm] \small \textbf{Tasas de inflación intertrimestral efectiva e interanual esperada}}
    \centering
    \includegraphics[scale=0.5]{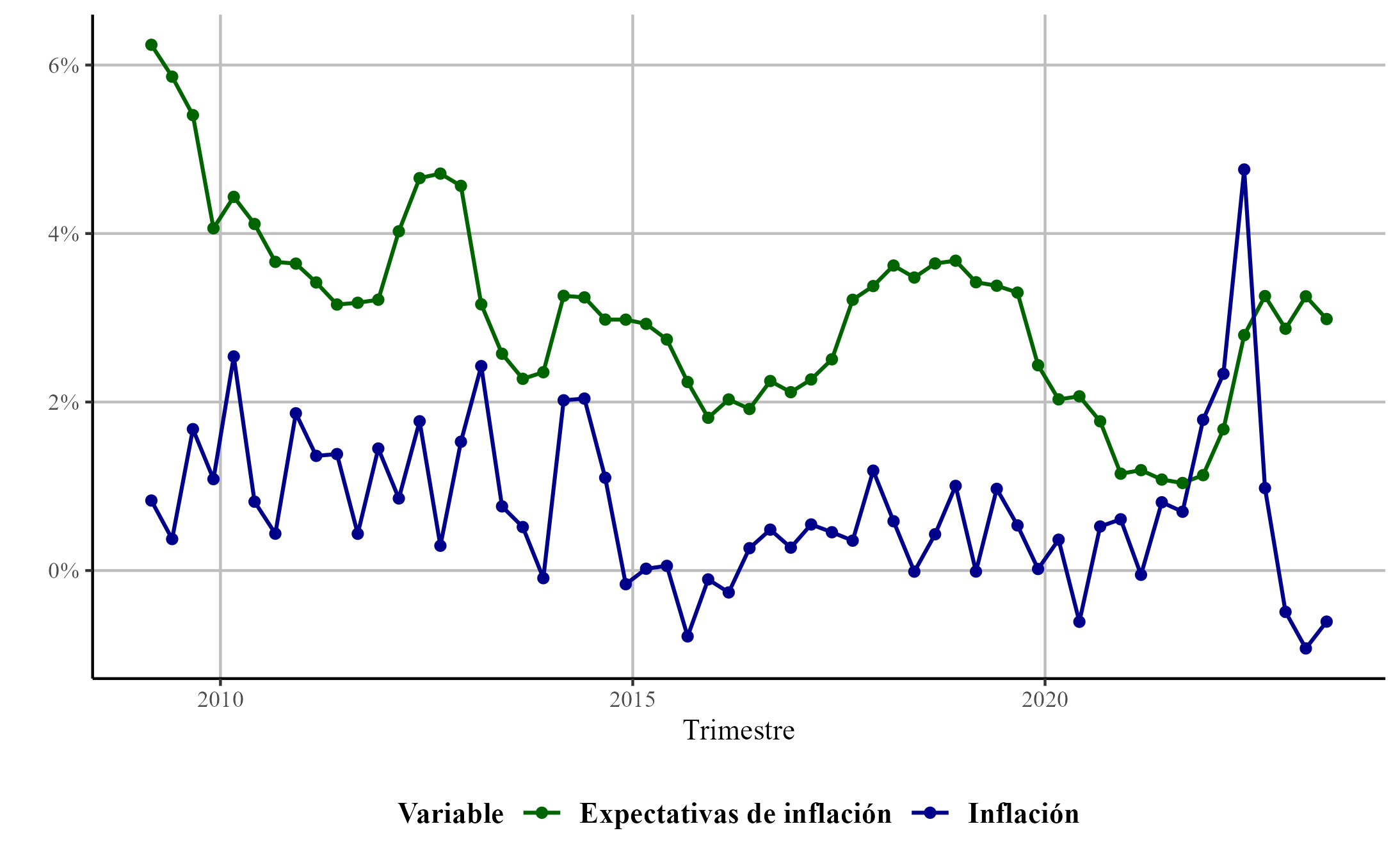}
    \begin{tablenotes}
\footnotesize
 \item \textbf{Fuente:} Elaboración propia .
    \end{tablenotes}
    \label{fig:plot_expect_mercado_q}
\end{figure}

  En la figura \ref{fig:plot_expect_mercado_q} se observa la tasa de inflaci\'on intertrimestral y las expectativas de inflaci\'on interanual. En general, se puede observar una tendencia decreciente de las expectativas de inflaci\'on y un comportamiento muy similar al de la inflaci\'on efectiva. 

\par
\begin{figure}[H]
\centering
    \caption{\\[0.0001cm] \small \textbf{Tasa de inflación intertrimestral y brecha del PIB trimestral}}
    \centering
    \includegraphics[scale=0.5]{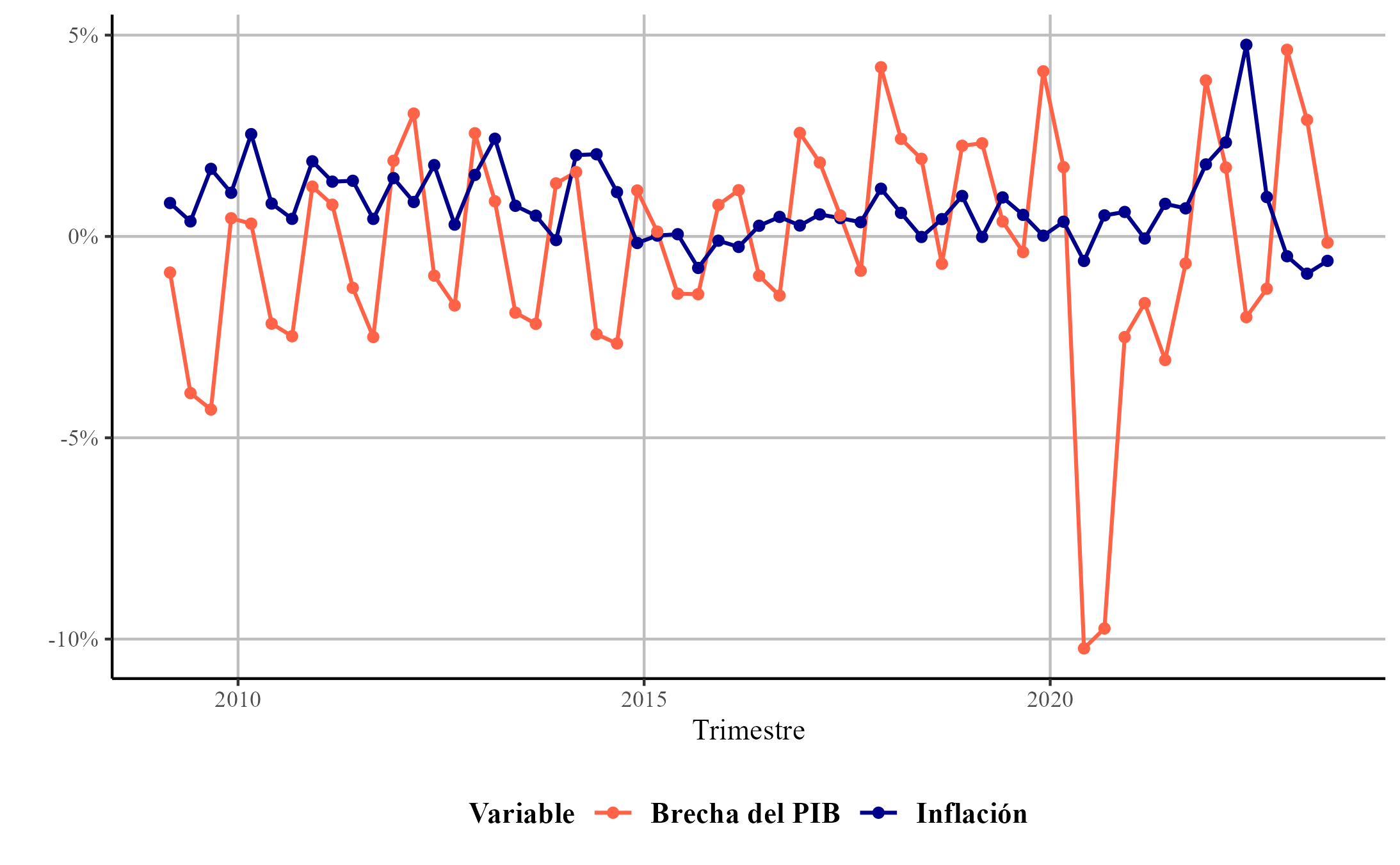}
    \begin{tablenotes}
\footnotesize
 \item \textbf{Fuente:} Elaboración propia .
    \end{tablenotes}
    \label{fig:plot_brecha_q}
\end{figure}

 En la figura \ref{fig:plot_brecha_q} se muestra la tasa de inflaci\'on y brecha del producto. Destacan tres observaciones. Primero, la brecha del producto parece tener un importante componente estacional con importantes subidas en el primer trimestre del a\~{n}o. Segundo, la tasa de inflaic\'on parece tener un comportamiento proc\'icliclo en el sentido de presentar variaciones en la misma direcci\'on que la brecha en el PIB, si bien estos cambios no se dan de forma proporcional, como tambi\'en lo sugieren las medidas de correlaci\'on ya vistas. Tercero, exceptuando las observaciones del 2020 y 2021, los trimestres con deflaci\'on han estado acompa\~{n}ados de brechas en el PIB no negativas o negativas pero relativamente cercanas a cero, lo que sugiere que las presiones deflacionarias a las que ha estado expuesto el pa\'is no provienen de la actividad econ\'omica real. 

\par
\begin{figure}[H]
\centering
    \caption{\\[0.0001cm] \small \textbf{Tasas de inflación intertrimestrales dom\'estica y de materias primas importadas}}
    \centering
    \includegraphics[scale=0.5]{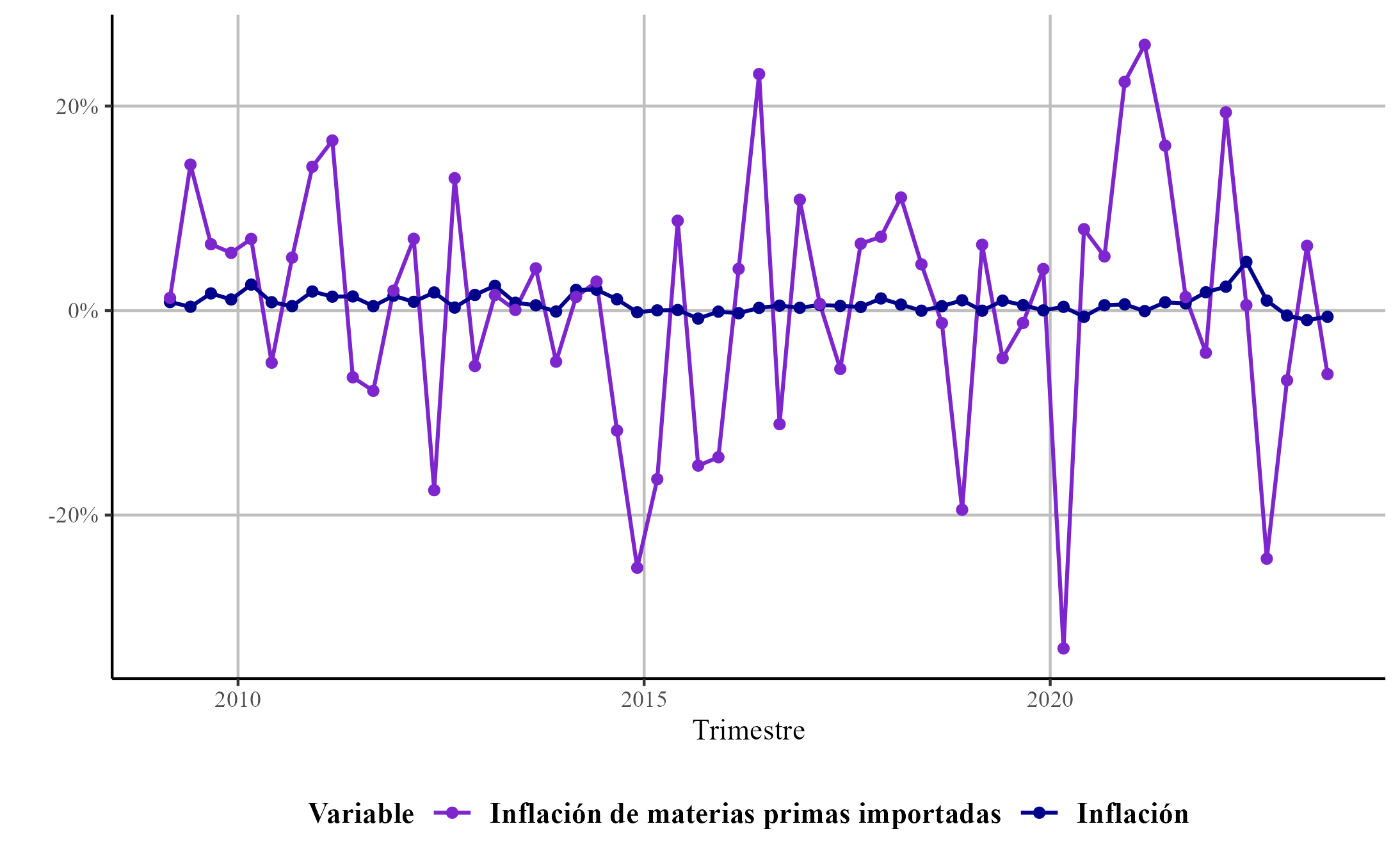}
    \begin{tablenotes}
\footnotesize
 \item \textbf{Fuente:} Elaboración propia .
    \end{tablenotes}
    \label{fig:plot_inf_imp_q.png}
\end{figure}

 Finalmente, en la figura \ref{fig:plot_inf_imp_q.png} se muestran las tasas de inflaci\'on dom\'estica y de materias primas. Para todo el periodo las materias primas importadas presentan una mayor volatilidad en el cambio de sus precios y presentan valores absolutos m\'as altos que la inflaci\'on dom\'estica. Se pueden observar fuertes periodos de deflaci\'on en las importaciones de materias primas alrededor del 2015 lo cual podr\'ia explicar la deflaci\'on observada en la econom\'ia dom\'estica. 

\section{Resultados}

\subsection{Resultados de estimaci\'on}
En este trabajo se estimaron modelos correspondientes a dos grupos de modelos para proyectar la inflaci\'on, a saber, univariados autoregresivos y multivariados. En el primer caso, se planete\'o estimar dos modelos ARIMA: uno para los rezagos sugeridos por las funciones de correlaci\'on completa y parcial y el otro elegidos por criterio AIC. En el primer caso se obtuvo un modelo ARMA(1,1), mientras que en el segundo un modelo ARMA(0,5) con intercepto; los valores AIC fueron, respectivamente, -351.3 y -347.06,  respectivamente. Dado ello, como \say{representante} de los modelos univariados, se utiliza el modelo ARMA(0,5), cuyos resultados de estimaci\'on son:  

\begin{center}
$    \begin{array}{cccccc}
        \pi_t&=0.3902 \epsilon_{t-1}+0.3409\epsilon_{t-2}+0.0514\epsilon_{t-3}-0.0011\epsilon_{t-4}+0.6150\epsilon_{t-5}\\
        &+0.0077
    \end{array}   $ 
\end{center}

Por otro lado, la elecci\'on de un modelo multivariado v\'ia mejores subconjuntos bajo criterio AIC arroj\'o un modelo con covariables dadas por expectativas de inflaci\'on, tercer y cuarto rezagos de la brecha del PIB y primer rezago en la inflaci\'on de materias primas importadas; los resultados se muestran en el cuadro \ref{estim_phil}. Todas las covariables excepto el segundo rezago en la brecha del PIB presentan un signo positivo. Las expectativas de inflaci\'on y la inflaci\'on de materias primas importadas parecen generar presiones al alza sobre la inflaci\'on dom\'estica, lo cual coincide con el actual consenso sobre el tema. Sin embargo, el segundo rezago en la brecha del PIB presenta un signo negativo lo cual puede ser problem\'atico. Las expectativas de inflaci\'on parecen ser el principal determinante de la inflaci\'on presente y es de notar que estas entraron en el modelo en su valor contempor\'aneo y no rezagado. 

\begin{table}[ht]
\centering
\centering
\caption{\\[0.0001cm] \small \textbf{Resultados de estimaci\'on, modelo multivariado}}
\begin{tabular}{rrrrr}
  \hline
 & Estimaci\'on & Error est\'andar & Valor t & Valor p \\ 
  \hline 
$E_t\{\pi^{yr}_{t+4}\}$ & 0.2462 & 0.0294 & 8.36 & 0.0000 \\  
  $Y^G_{t-2}$ & -0.0443 & 0.033 & -1.34 & 0.187 \\  
  $\pi^{Imports}_{t-1}$ & 0.0188 & 0.0078 & 2.41 & 0.02 \\ 
   \hline
\end{tabular}
\label{estim_phil}
\end{table}

De acuero con lo establecido en la secci\'on \ref{metod}, se estim\'o un modelo ARIMA para las expectativas de inflaci\'on, brecha en el PIB e inflaci\'on de materias primas importadas. Para esta \'ultima se obtuvo un un modelo ARMA(0,0) lo que indica que esta constituye un ruido blanco. Para las dos primeras se obtuvo:

\begin{align*}
    E_t\{\pi^{yr}_{t+4}\} &=1.2631 E_{t-1}\{\pi^{yr}_{t+3}\} -0.3435E_{t-2}\{\pi^{yr}_{t+2}\} +0.0311 \\[0.3cm]
    Y^G_t&=0.744Y^G_{t-1}-0.4366Y^G_{t-2}+0.1429Y^G_{t-3}\\
    &+0.3565Y^G_{t-4}-0.5029Y^G_{t-5}  
\end{align*}

\subsection{Evaluaci\'on del desempe\~{n}o predictivo}
\begin{figure}[H]
\centering
\caption{\\[0.0001cm] \small \textbf{Tasa de inflaci\'on ajustada y proyectada}}
\begin{subfigure}[b]{0.45\textwidth}
    \centering
    \includegraphics[width=\textwidth]{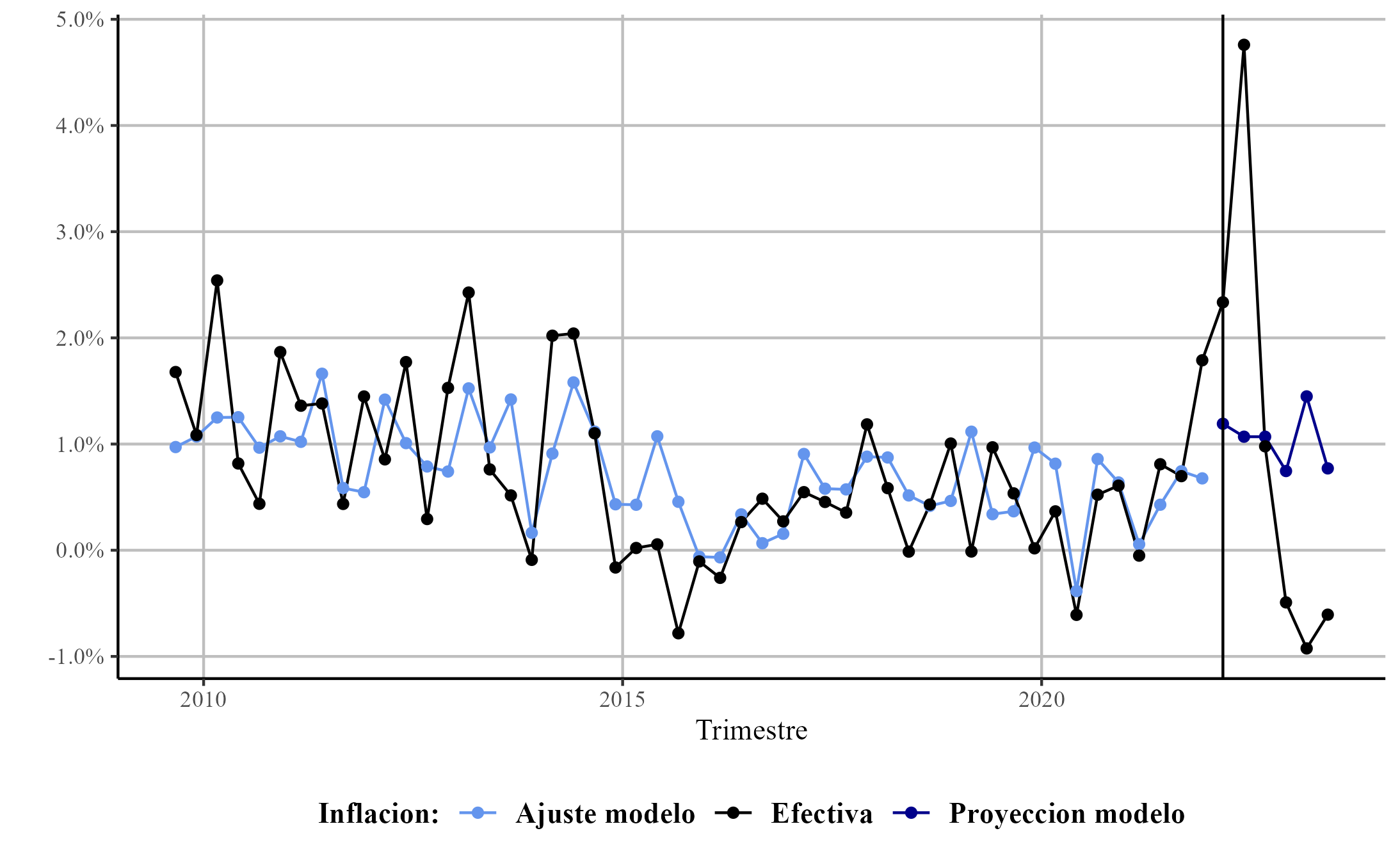}
    \caption{Modelo univariado}
\end{subfigure}
\hfill
\begin{subfigure} [b]{0.45\textwidth}
    \centering
    \includegraphics[width=\textwidth]{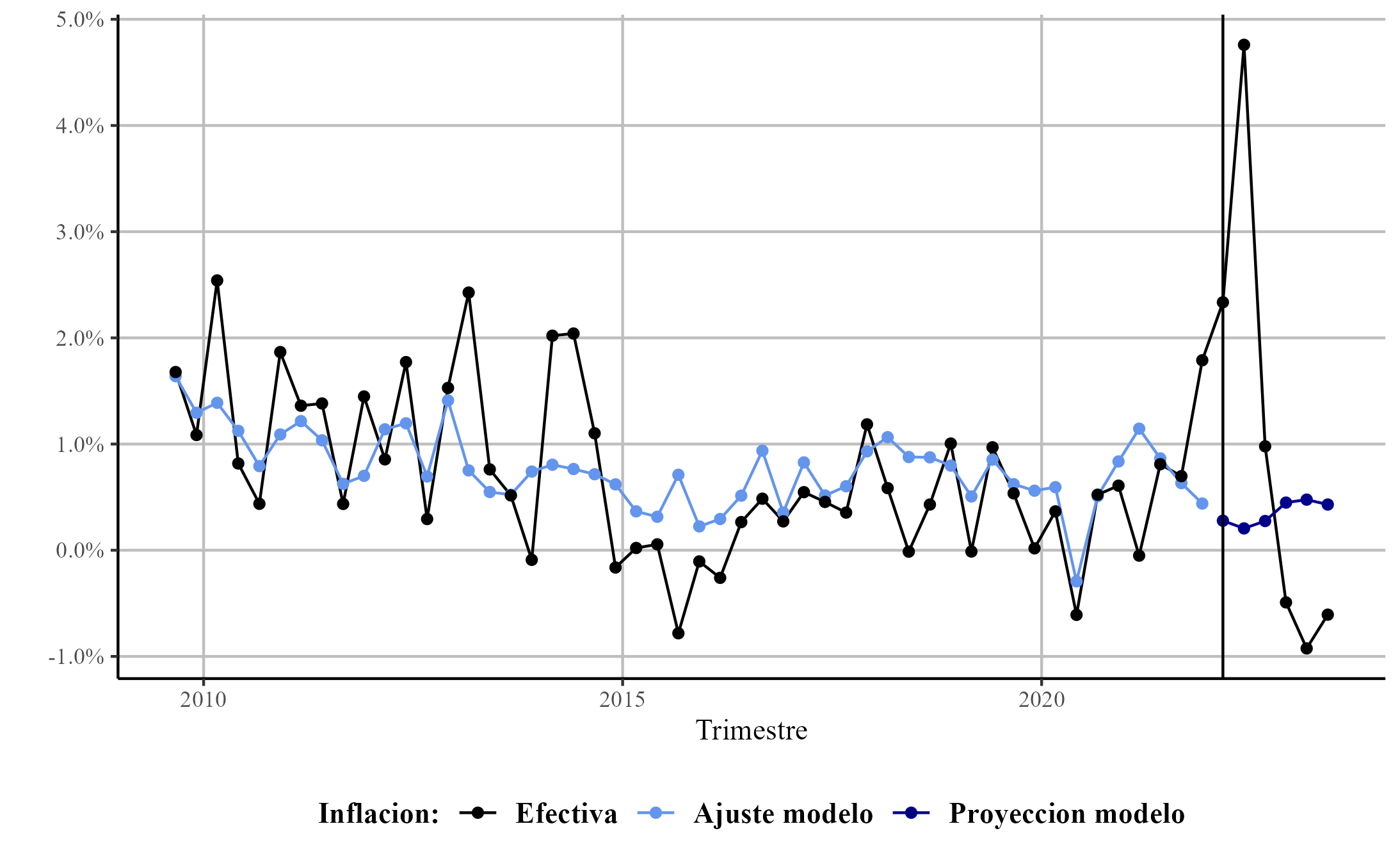}
    \caption{Modelo multivariado}
\end{subfigure}
\label{fig:mods_pred}
\end{figure}

En esta secci\'on se eval\'ua el desempe\~{n}o predictivo de ambos modelos. En la figura \ref{fig:mods_pred} se muestran las proyecciones y ajustes para cada modelo\footnote{El ajuste se tiene previo a la fecha quiebre que se representa con la recta vertical y la proyecci\'on para las fechas posteriores a esta.}. En estas se pueden observar tres principales hallazgos. Primero, el modelo multivariado parece tener un desempe\~{n}o pobre fuera de la muestra; en particular, no es capaz de predecir los shocks inflacionarios ocurridos al inicio del 2022 provocados por la reapertura del las econom\'ias tras la pandemia y el conflicto en Ucrania y siempre predice un movimiento en la direcci\'on contraria a la observada. Esto puede deberse a que la brecha del PIB entra con un signo negativo en el modelo, lo cual contradice a la postulaci\'on te\'orica y podr\'ia deberse al m\'etodo con el que se estim\'o el PIB potencial. Segundo, el modelo univariado parece tener una capacidad predictiva ligeramente mejor fuera de la muestra, pero a\'un as\'i no muy bueno. Tercero, en el modelo multivariado la proyecci\'on parece tener un importante componente c\'iclico que podr\'ia ser generado por las estimaciones autoregresivas de sus covariables, particularmente la brecha del producto. Es de crucial importancia tener en cuenta que esta an\'alisis fuera de muestra se da en un periodo de condiciones econ\'omicas mundiales inusuales, por lo que su alcance es limitado, si bien s\'i permite evaluar la capacidad de predicci\'on de estos modelos \emph{bajo tal tipo de condiciones}. Como es usual en la literatura, se encuentra que ninguno de los dos ac\'a estimados son robustos ante \emph{shocks} del tipo vividos en el 2022. Idealmente, el modelo mutivariado deber\'ia ser capaz de incorporar el efecto de tales perturbaciones a trav\'es de sus covariables. Sin embargo, en este caso, el modelo multivariado no parece lograrlo, lo cual puede deberse al hecho de que las predicciones de este se basaron, a su vez, en predicciones de las covariables que fueron obtenidas de modelos univariados.

 En el cuadro \ref{cuad_contraste} varias medidas del desempe\~{n}o y capacidad predictiva de los dos modelos. El modelo univariado presenta menores RMSE y MSE fuera de muestra, sin embargo, estos son muy cercanos a los del modelo multivariado, por lo que apenas sugiere, a lo sumo, una capacidad marginalmente mayor del modelo univariado. Estas mismas medidas calculadas dentro de la muestra son, al contrario, menores en el modelo multivariado, lo que sugiere un mejor ajuste del mismo. Por otro lado, el modelo multivariado (i.e. la regresi\'on multivariada), presenta un AIC significativamente menor que el univariado, lo que sugiere una mayor capacidad predictiva de este, si bien esta m\'etrica no toma en cuenta la capacidad de las regresiones univariadas que se utilizan para alimentarle al momento de proyectar. De nuevo, es de notar que, si bien el modelo multivariado ac\'a utilizado permite incluir din\'amicas inflacionarias m\'as complejas dadas, en las predicciones de este subyacen las proyecciones de las covariables, las cuales se realizaron con modelos univariados. Por ello, quiz\'as, la capacidad predictiva del modelo multivariado podr\'ia aumentarse de forma importante si m\'as bien se proyectasen las covariables con modelos multivariados m\'as complejos que permitan recoger mayores din\'amicas; e.g. un modelo VAR. \\
\
\begin{table}[ht]
\centering
\caption{Medidas de capacidad predictivda de los modelos}
\begin{tabular}{rlrr}
  \hline
  Medida & Modelo univariado & Modelo multivariado \\ 
  \hline
    RMSE fuera de muestra  & 0.020092 & 0.022132 \\ 
    MAE fuera de muestra & 0.016685 & 0.0180580 \\ 
    AIC intramuestral & -334.617 & -341.11506 \\ 
    RMSE intramuestral & 0.000053 & 0.000047 \\
    MAE intramuestral & 0.00602 & 0.00497 \\
   \hline 
\end{tabular}
\label{cuad_contraste}
\end{table}

\section{Limitaciones}
Es necesario tener en cuenta varias de las limitaciones ya mencionadas, tanto del modelo como de la evaluaci\'on de ambos. Primero, en la proyecci\'on con el modelo multivariado, la cual permite incluir din\'amicas inflacionarias m\'as complejas, subyacen las proyecciones de sus covariables v\'ia modelos univariados. Por ello, puede heredar cierta inflexibilidad de los modelos univariados ante \emph{shocks}. Ello es de particular importancia en la medida en la que limita la capacidad del modelo multivariado de contemplar los \emph{shocks} vividos a inicio del 2022. Segundo, la evaluaci\'on del desempe\~{n}o fuera de muestra se dio en un periodo de fuertes \emph{shocks} en las econom\'ias internacionales, lo cual, si bien permite evaluarles en tales escenario, plantea un desaf\'io quiz\'as hiperambicioso para el tiempo de modelos ac\'a estimados. Tercero,  

\section{Conclusiones}
En este trabajo se estimaron dos modelos para predecir la tasa de inflaci\'on, uno correspondiente al grupo de modelos univariados autoregresivos y el otro a los modelos multivariados a fin de contrastar su desmepe\~{n}o. Lo encontrado sugiere que el modelo univariado posee una capacidad predictiva marginalmente fuera de la muestra, mientras que el univariado lo tiene previo a la muestra. Ambos modelos presentan un desmepe\~{n}o no muy bueno fuera de la muestra, si bien esto es de esperar ya que en tal periodo de tiempo se dieron shocks importantes en las econom\'ias. 
\newpage
\bibliography{bibliografia}
\newpage
\section{Anexos}
Ac\'a se meustran las pruebas de estacionariedad y de supuestos de los mdoelos de regresi\'on utilizados. 
\subsection{Pruebas de estacionariedad}
\begin{table}[h]
\centering
\centering
\caption{\\[0.0001cm] \small \textbf{P valor de las pruebas de estacionariedad para la inflación del IPC }}
\begin{tabular}{rrr}
  \hline
  Modelo & Rezago con información & Rezago=4  \\ 
  \hline 
Deriva & 0.0000** & 0.0000**  \\  
  Tendencia & 0.0000** & 0.0000**  \\ 
   \hline
\end{tabular}
\label{estim_phil}
\end{table}

\begin{table}[H]
\centering
\centering
\caption{\\[0.0001cm] \small \textbf{P valor de las pruebas de estacionariedad para la inflación de las importaciones de materias primas}}
\begin{tabular}{rrr}
  \hline
  Modelo & Rezago con información & Rezago=4  \\ 
  \hline 
Deriva & 0.0000** & 0.0000**  \\  
  Tendencia & 0.0000** & 0.0000**  \\ 
   \hline
\end{tabular}
\label{estim_phil}
\end{table}

\begin{table}[H]
\centering
\centering
\caption{\\[0.0001cm] \small \textbf{P valor de las pruebas de estacionariedad para las expectativas de mercado}}
\begin{tabular}{rrr}
  \hline
  Modelo & Rezago con información & Rezago=4  \\ 
  \hline 
Deriva & 0.0000** & 0.0000**  \\  
  Tendencia & 0.0000** & 0.0000**  \\ 
   \hline
\end{tabular}
\label{estim_phil}
\end{table}

\begin{table}[H]
\centering
\centering
\caption{\\[0.0001cm] \small \textbf{P valor de las pruebas de estacionariedad para el tipo de cambio}}
\begin{tabular}{rrr}
  \hline
  Modelo & Rezago con información & Rezago=4  \\ 
  \hline 
Deriva & 2.4745e-09 & 2.63e-09  \\  
  Tendencia & 0.000** & 0.0000**  \\ 
   \hline
\end{tabular}
\label{estim_phil}
\end{table}

\begin{table}[H]
\centering
\centering
\caption{\\[0.0001cm] \small \textbf{P valor de las pruebas de estacionariedad para la brecha del PIB}}
\begin{tabular}{rrr}
  \hline
  Modelo & Rezago con información & Rezago=4  \\ 
  \hline 
Deriva & 0.0000** & 0.0000**  \\  
  Tendencia & 0.0000** & 0.0000**  \\ 
   \hline
\end{tabular}
\label{estim_phil}
\end{table}

\newpage
\subsection{Validaci\'on de modelos}

\subsubsection{Modelo univariado}

\begin{table}[ht]
\centering
\caption{\\[0.0001cm] \small \textbf{P valor de las pruebas de estacionariedad para el tipo de cambio}}
\begin{tabular}{rrrlrlrr}
  \hline
 & rezagos & LjungBox & autocorr & white & heteros & jarque & shapiro \\ 
  \hline
1 &   1 & 0.86 & NO HAY & 0.99 & NO HAY & 0.27 & 0.05 \\ 
  2 &   2 & 0.96 & NO HAY & 0.32 & NO HAY & 0.27 & 0.05 \\ 
  3 &   3 & 0.99 & NO HAY & 0.17 & NO HAY & 0.27 & 0.05 \\ 
  4 &   4 & 0.75 & NO HAY & 0.88 & NO HAY & 0.27 & 0.05 \\ 
  5 &   5 & 0.37 & NO HAY & 0.98 & NO HAY & 0.27 & 0.05 \\ 
  6 &   6 & 0.40 & NO HAY & 0.42 & NO HAY & 0.27 & 0.05 \\ 
  7 &   7 & 0.48 & NO HAY & 0.49 & NO HAY & 0.27 & 0.05 \\ 
  8 &   8 & 0.59 & NO HAY & 0.07 & NO HAY & 0.27 & 0.05 \\ 
   \hline
\end{tabular}
\end{table}

\subsubsection{Modelo multivariado}

\begin{table}[ht]
\centering
\caption{\\[0.0001cm] \small \textbf{Resultados de diagn\'osticos de supuestos de regresi\'on multivariada}}
\begin{tabular}{rrrlrlrr} 
  \hline
 & rezagos & LjungBox & autocorr & white & heteros & jarque & shapiro \\ 
  \hline
1 &   1 & 0.35 & NO HAY & 0.68 & NO HAY & 0.08 & 0.01 \\ 
  2 &   2 & 0.65 & NO HAY & 0.41 & NO HAY & 0.08 & 0.01 \\ 
  3 &   3 & 0.82 & NO HAY & 0.14 & NO HAY & 0.08 & 0.01 \\ 
  4 &   4 & 0.51 & NO HAY & 0.62 & NO HAY & 0.08 & 0.01 \\ 
  5 &   5 & 0.38 & NO HAY & 0.45 & NO HAY & 0.08 & 0.01 \\ 
  6 &   6 & 0.48 & NO HAY & 0.68 & NO HAY & 0.08 & 0.01 \\ 
  7 &   7 & 0.56 & NO HAY & 0.59 & NO HAY & 0.08 & 0.01 \\ 
  8 &   8 & 0.66 & NO HAY & 0.46 & NO HAY & 0.08 & 0.01 \\ 
   \hline
\end{tabular}
\end{table}

\begin{table}[ht]
\centering
\begin{tabular}{rrrlrlrr}
  \hline
 & rezagos & LjungBox & autocorr & white & heteros & jarque & shapiro \\ 
  \hline
1 &   1 & 0.63 & NO HAY & 0.68 & NO HAY & 0.90 & 0.83 \\ 
  2 &   2 & 0.55 & NO HAY & 0.27 & NO HAY & 0.90 & 0.83 \\ 
  3 &   3 & 0.69 & NO HAY & 0.33 & NO HAY & 0.90 & 0.83 \\ 
  4 &   4 & 0.59 & NO HAY & 0.41 & NO HAY & 0.90 & 0.83 \\ 
  5 &   5 & 0.72 & NO HAY & 0.55 & NO HAY & 0.90 & 0.83 \\ 
  6 &   6 & 0.81 & NO HAY & 0.99 & NO HAY & 0.90 & 0.83 \\ 
  7 &   7 & 0.89 & NO HAY & 0.13 & NO HAY & 0.90 & 0.83 \\ 
  8 &   8 & 0.92 & NO HAY & 0.76 & NO HAY & 0.90 & 0.83 \\ 
   \hline
\end{tabular}
\end{table}
\begin{table}[ht]
\centering
\begin{tabular}{rrrlrlrr}
  \hline
 & rezagos & LjungBox & autocorr & white & heteros & jarque & shapiro \\ 
  \hline
1 &   1 & 0.26 & NO HAY & 0.79 & NO HAY & 0.91 & 0.77 \\ 
  2 &   2 & 0.39 & NO HAY & 0.15 & NO HAY & 0.91 & 0.77 \\ 
  3 &   3 & 0.20 & NO HAY & 0.11 & NO HAY & 0.91 & 0.77 \\ 
  4 &   4 & 0.08 & NO HAY & 0.78 & NO HAY & 0.91 & 0.77 \\ 
  5 &   5 & 0.11 & NO HAY & 0.12 & NO HAY & 0.91 & 0.77 \\ 
  6 &   6 & 0.10 & NO HAY & 0.78 & NO HAY & 0.91 & 0.77 \\ 
  7 &   7 & 0.04 & Si hay & 0.10 & NO HAY & 0.91 & 0.77 \\ 
  8 &   8 & 0.05 & Si hay & 0.85 & NO HAY & 0.91 & 0.77 \\ 
   \hline
\end{tabular}
\end{table}
\begin{table}[ht]
\centering
\begin{tabular}{rrrlrlrr}
  \hline
 & rezagos & LjungBox & autocorr & white & heteros & jarque & shapiro \\ 
  \hline
1 &   1 & 0.72 & NO HAY & 0.50 & NO HAY & 0.65 & 0.61 \\ 
  2 &   2 & 0.83 & NO HAY & 0.81 & NO HAY & 0.65 & 0.61 \\ 
  3 &   3 & 0.92 & NO HAY & 0.09 & NO HAY & 0.65 & 0.61 \\ 
  4 &   4 & 0.64 & NO HAY & 0.07 & NO HAY & 0.65 & 0.61 \\ 
  5 &   5 & 0.55 & NO HAY & 0.04 & Si hay & 0.65 & 0.61 \\ 
  6 &   6 & 0.68 & NO HAY & 0.42 & NO HAY & 0.65 & 0.61 \\ 
  7 &   7 & 0.78 & NO HAY & 0.60 & NO HAY & 0.65 & 0.61 \\ 
  8 &   8 & 0.82 & NO HAY & 0.17 & NO HAY & 0.65 & 0.61 \\ 
   \hline
\end{tabular}
\end{table}
\end{document}